\input{aipcheck}

\documentclass[
    ,final            
  ,numberedheadings 
  ]
  {aipproc}

\layoutstyle{6x9}

\def\to{\rightarrow}

\def\bi{\begin{itemize}}
\def\ei{\end{itemize}}

\def\ta{\tilde a}

\def\tb{\tilde b}

\def\tst{\tilde t}
\def\ttau{\tilde \tau}

\def\tg{\tilde g}

\def\tw{\widetilde W}
\def\tz{\widetilde Z}

\def\agt{\stackrel{>}{\sim}}
\def\be{\begin{equation}}  
\def\ee{\end{equation}}  

\newcommand\prd[3]{{\it Phys.\ Rev.\ }{\bf D #1} (#2) #3}

\newcommand\prl[3]{{\it Phys.\ Rev.\ Lett.\ }{\bf #1} (#2) #3}
\newcommand\plb[3]{{\it Phys.\ Lett.\ }{\bf B #1} (#2) #3}

\newcommand\jhep[3]{{\it J. High Energy Phys.\ }{\bf #1} (#2) #3}

\newcommand\npb[3]{{\it Nucl.\ Phys.\ }{\bf B #1} (#2) #3}

\newcommand{\hepph}[1]{hep-ph/#1}

\begin{document}

\title{Dark matter from SUGRA GUTs: mSUGRA, NUSUGRA and
Yukawa-unified SUGRA}


\author{Howard Baer}{
  address={Dep't of Physics and Astronomy, University of Oklahoma,
Norman, OK 73019}
}

\begin{abstract}
Gravity-mediated SUSY breaking models with $R$-parity 
conservation give rise to dark matter in the universe.
I review neutralino dark matter in the minimal
supergravity model (mSUGRA), models with non-universal
soft SUSY breaking terms (NUSUGRA) which yield a 
well-tempered neutralino, and models with unified 
Yukawa couplings at the GUT scale (as may occur in an $SO(10)$ SUSY GUT
theory). These latter models have difficulty 
accommodating neutralino dark matter, but work very well if the dark matter
particles are axions and axinos.

\end{abstract}

\maketitle


This talk covers several highlights from research into supersymmetric 
dark matter by our research group over the past several years.

\section{Neutralino dark matter in the mSUGRA model}

The minimal supergravity model\cite{msugra}, 
mSUGRA or CMSSM, is the paradigm
model for many investigations of supersymmetry phenomenology.
The mSUGRA model has the MSSM embedded in a supergravity framework, 
and then arranges for supergravity breaking via the super-Higgs mechanism,
in a so-called hidden sector of the theory. Upon supergravity breaking, 
the gravitino acquires a mass of order $m_{3/2}\sim M^2/M_{Pl}\sim 1$ TeV,
so that $M\sim 10^{11}$ GeV. Soft SUSY breaking (SSB) terms are induced
due to the breakdown of supergravity, leading to weak scale SSB
masses for gauginos, scalars, trilinear and bilinear soft terms.
The defining assumption for mSUGRA is that at the GUT scale 
all scalars receive a common mass $m_0$, all gauginos receive a common 
mass $m_{1/2}$, $A$-terms receive a common mass $A_0$ and a bilinear 
mass term $B$ is also induced. The SSB terms evolve from $M_{GUT}$ 
down to $M_{weak}$ according to the renormalization group 
equations (RGEs). In particular, the up-Higgs soft mass 
$m_{H_{u}}^2$ is driven to negative values by the large top quark 
Yukawa coupling, causing a breakdown in electroweak symmetry (EWSB).
The scalar potential minimization conditions allow one to trade 
the parameter $B$ for $\tan\beta$, the ratio of Higgs field vevs, 
while the magnitude (but not the sign) of the superpotential
Higgs mass term $\mu$ is fixed in terms of the measured value of $M_Z$.
The well-known parameter space
\begin{equation}
m_0,\ m_{1/2},\ A_0,\ \tan\beta ,\ sign(\mu )
\end{equation}
allows one to calculate all sparticle masses, mixings, scattering 
cross sections, decay rates and, in the case of the
lightest neutralino $\tz_1$ (assumed to be the stable lightest
SUSY particle (LSP), and a good WIMP dark matter candidate), the
relic dark matter abundance\cite{wss}.

\begin{figure}
  \includegraphics[height=.3\textheight]{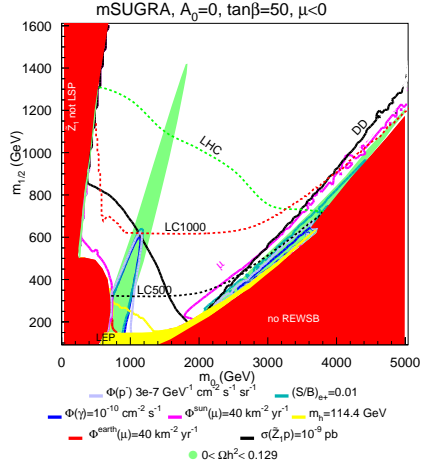}
  \caption{Expose of allowed regions of the mSUGRA model
$m_0\ vs.\ m_{1/2}$ plane for $\tan\beta =50$, $A_)=0$ and 
$\mu <0$. We show dark matter allowed regions (green), 
collider reaches, and DD and IDD reach contours.}
\label{fig:msugra}
\end{figure}

The WMAP collaboration\cite{wmap}, and other groups, have measured the 
dark matter abundance of the universe to be $\Omega_{CDM}h^2\simeq 0.11$,
which highly constrains models of new physics containing dark matter candidates,
and in this case the mSUGRA model. 
In Fig. \ref{fig:msugra}, the $m_0\ vs.\ m_{1/2}$ plane of
mSUGRA parameter space is shown for $\tan\beta =50$, $A_0=0$ and
$\mu <0$. We use isajet for our spartciel mass computations\cite{isajet}.
The red regions are not allowed due to (left-edge)
the presence of a charged, stable stau $\ttau_1$ LSP
(in conflict with negative searches for charged/colored relic 
from the Big Bang), or lack of appropriate breakdown of electroweak 
symmetry (lower-right red region). The green-shaded regions give $\Omega_{\tz_1}h^2\le 0.13$, 
and so are in accord with WMAP measurements; the white-shaded regions give
$\Omega_{\tz_1}h^2>0.13$ and so are presumably excluded by the measured dark matter abundance.

The dark matter allowed regions consist of:
\begin{itemize}
\item A bulk region at low $m_0$ and low $m_{1/2}$ where neutralinos annihilate
via $t$-channel slepton exchange (covered over here by red because we are at such high $\tan\beta$).
\item The stau co-annihilation region where $\tz_1 -\ttau_1$ can co-annihilate due to their 
small mass gap (very thin sliver adjacent to red region at low $m_0$. 
\item The focus point (FP) region at large $m_0$ where $\mu$ becomes small and $\tz_1$ becomes
mixed bino-higgsino dark matter.
\item The $A$-annihilation funnel in the middle of the plot where $\tz_1\tz_1$
annihilation through the $A$-resonance is enhanced because $2m_{\tz_1}\simeq m_A$.
\item There is also a stop co-annihilation region for specific $A_0$ values, and
a light Higgs $h$-resonance annihilation region possible at low $m_{1/2}$ for
lower $\tan\beta$ values.
\end{itemize}

We also super-pose on the plot the approximate reach of the Fermilab Tevatron
via clean trilepton channel for 10 fb$^{-1}$ of integrated luminosity, and the LHC
reach for 100 fb$^{-1}$ of integrated luminosity. The LHC reach\cite{lhc} covers essentially all
the stau co-annihilation region and most of the $A$-funnel, but a long strip of FP 
region extends away from the LHC reach where gluinos and squarks are very heavy, but charginos 
are quite light and higgsino-like, since $\mu$ is small. Notice in this
region the reach of a linear $e^+e^-$ collider can exceed that of LHC, since chargino pair production
is easy to see at linear colliders, but hard to see at LHC\cite{ilc}.

We also show contours of direct dark matter detection (DD), for experiments such as CDMS, Xenon-100,
LUX or WARP (black contour). Note that this contour covers the FP region, so if SUSY lies in
the FP region, with $m_{1/2}\agt 700$ GeV, then DD experiments will soon find a signal, 
while LHC may see none!

The magenta contour labelled $\mu$ denotes the approximate reach of the IceCube
neutrino detector: it also covers most of the FP region. 
We also show approximate reach contours for indirect dark matter detection (IDD)
via positrons, anti-protons and $\gamma$-rays arising from WIMP annihilation in 
the galactic halo\cite{bbko}. These contours fill much of the $A$-funnel and also the FP region. 
In the case of the $A$-funnel, halo neutralinos have an enhanced annihilation rate through the
$A$-resonance\cite{bo}, while halo annihilations through $h$ and $H$ are suppressed, 
since $\sigma\cdot v\to 0$ as the WIMP velocity $v\to 0$.

The three main regions of mSUGRA parameter space can be characterized by their 
direct and indirect detection rates:
1. For co-annihilation regions, one expects halo annihilation rates to be small,
since co-annihilation cannot take place in the galactic halo. Also, 
DD and $\nu$-telescope rates may be very small. 2. In the
$A$-annihilation funnel, halo annihilation rates can be large, but DD and 
$\nu$-telescope rates can be small.  3. In the FP region,
all of halo annihilation, DD and $\nu$-telescope detection rates can be large.

\section{Well-tempered neutralinos in SUGRA models
with non-universality}

While mSUGRA may be the most popular model for many SUSY analyses, there is strong motivation
for SUGRA models with non-universality.
A simple example ocurs in $SO(10)$ SUSY GUTs. Here, the Higgs multiplets may occupy the 
fundamental $\bf 10$ of $SO(10)$, while matter scalars occupy the spinorial $\bf 16$: one would 
expect in general $m_{10}^2\ne m_{16}^2$. With this single additional parameter, 
for any point in mSUGRA parameter space with too large $\Omega_{\tz_1}h^2$, one might dial\cite{nuhm1}
$m_{10}>m_{16}$ and reach mixed higgsino dark matter (even though one is not in the FP region),
or one may dial $m_{10}^2$ to negative values and enter the $A$-funnel (even at low $\tan\beta$).
The first of these situations is an example of a ``well-tempered neutralino'', wherein its
composition is adjusted to gain the measured relic density\cite{wtn1}. 
When one gives it enough higgsino
component to gain the measured relic density, one also increases the direct and 
indirect detection rates\cite{wtn2}.

\begin{figure}
  \includegraphics[height=.3\textheight]{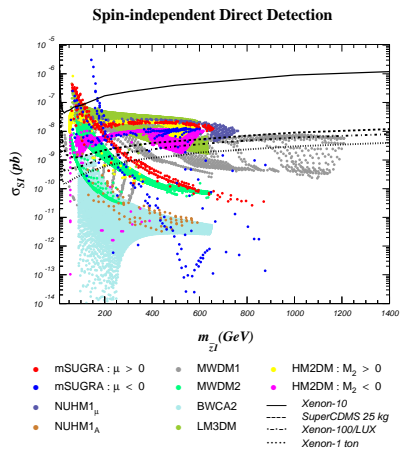}
  \caption{Direct detecion rates for a variety of models with 
well-tempered neutralinos. Note the asymptotic regions
extending across $10^{-8}$ pb!}
\label{fig:wtn}
\end{figure}

A variety of models with well-tempered neutralinos are shown in Fig. \ref{fig:wtn}. There are
one-parameter non-universal Higgs models ($NUHM_\mu$ and $NUHM_A$)\cite{nuhm1}, models with 
mixed wino-bino-higgsino dark matter (MWDM1 and MWDM2)\cite{mwdm}, high $M_2$
mixed bino-higsino dark matter (HM2DM)\cite{hm2dm}, low $M_3$ mixed bino-higgsino dark matter
(LM3DM)\cite{lm3dm} and pure bino dark matter with its mass (not composition) tempered to 
allow for bino-wino co-annihilation (BWCA)\cite{bwca}.
Note, parameters are dialed so that every point has exactly the right relic density
$\Omega_{\tz_1}h^2\simeq 0.11$. A few models with mass-tempering (such as BWCA) have low
direct detection rates, but the models with neutralino {\it composition} tempering form
an asymptote at $\sigma_{SI}(\tz_1 p)\sim 10^{-8}$ pb. This cross section should be accessible to
Xenon-100, LUX, WARP-140 and superCDMS. Thus, once the {\it special} $10^{-8}$ pb SI scattering cross-section is
well explored, either well-tempered neutralino dark matter will be discovered, or this whole class 
of models will be excluded\cite{wtn2}! The FP region of mSUGRA of course falls in this region as well.

\section{Mixed axion/axino dark matter in Yukawa-unified models}

One of the great successes of $SU(5)$ GUT theories was the prediction of
$b-\tau$ Yukawa coupling unification. In the simplest $SO(10)$ SUSY GUT
models, one expects the more restrictive condition of $t-b-\tau$
Yukawa coupling unification. It was recognized very early on that
one criteria for this to occur in the MSSM is that $\tan\beta$
needs to be very large: $\tan\beta \sim 50$. At these high values of
$\tan\beta$, SM-MSSM threshold corrections to the $b$-quark Yukawa coupling
become very large\cite{hrs}. These occur mainly through $\tst_i\tw_j$ and 
$\tg\tb_i$ loop diagrams. The threshold corrections thus cause the entire 
calculation to depend on the spectrum of SUSY particles.

We assumed that the MSSM was the correct effective field theory between $M_{weak}$
and $M_{GUT}$, but that the parameter space at $M_{GUT}$ was that of $SO(10)$:
\begin{equation}
m_{16},\ m_{10},\ M_D^2,\ m_{1/2},\ A_0,\ \tan\beta,\ {\rm and}\ sign(\mu ) .
\end{equation}
Here $M_D$ parametrizes the splitting of the Higgs SSB terms and other scalars:
$m_{H_{u,d}}^2=m_{10}^2\mp 2M_D^2$. Yukawa coupling unification succeeds best when the
$D$-term splitting is only applied to the Higgs scalars, and not other matter scalars\cite{bdr}.

\begin{figure}
  \includegraphics[height=.4\textheight]{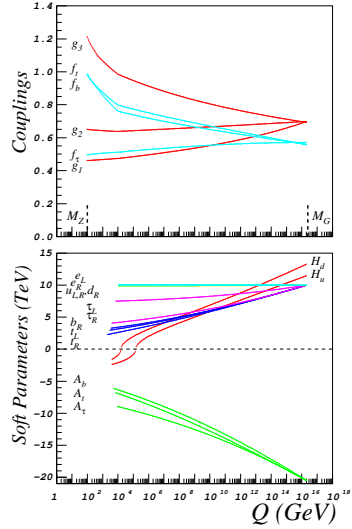}
  \caption{Evolution of gauge and Yukawa couplings
(upper frame) and SSB terms (lower frame) in Yukawa-unified
SUSY models.}
\label{fig:so10}
\end{figure}

A scan over parameter space, using the Isajet/Isasugra spectrum generator
(including full 2-loop RGE running and complete 1-loop sparticle mass and
Yukawa threshold corrections) finds that Yukawa unified solutions can in fact be found.
They are found for only very special choices of $SO(10)$ parameter choices:
1. $\tan\beta \sim 50$, 2. $m_{16}\sim 10$ TeV, while $m_{1/2}$ is very small,
3. $A_0^2=2m_{10}^2=4m_{16}^2$, 4. $\mu >0$, and 5. split Higgs mass at the GUT scale, with
$m_{H_u}<m_{H_d}$. The latter criteria is need for an appropriate breakdown of EW symmetry.
An example is given in Fig. \ref{fig:so10}. 

The spectrum of SUSY particle generated for Yukawa unified SUSY is also very unique:
1. first/second generation matter scalars around 10 TeV, 2. third generation scalars around a few TeV
due to the radiatively generated inverted scalar mass hierarchy\cite{bfpz},
3. gluino mass around $350-500$ GeV with charginos around $100-160$ GeV and 
$m_{\tz_1}\sim 50-80$ GeV\cite{bkss}. 

Since $\mu$ and $m_A$ are typically a few TeV, the 
$\tz_1$ is nearly pure bino-like. The heavy scalars imply that the neutralino relic density
$\Omega_{\tz_1}h^2$ is in the $10^1-10^4$ range: many orders of magnitude above its measured
value. An elegant way to solve this Yukawa-unified dark matter problem is to assume
a Peccei-Quinn solution to the strong $CP$ problem, which then implies that
a mixture of axions $a$  and axinos $\ta$ would actually constitute the dark matter. The $\tz_1$
lives a lifetime of about 1 sec, and decays before it can interfere with BBN:
$\tz_1\to\gamma\ta$. From supergravity theory, we expect the gravitino mass
$m_{3/2}\sim m_{16}\sim 10$ TeV. This is actually very propitious, as it allows for 
a solution of the BBN/gravitino problem and allows for a reheat temperature of the universe 
$T_R$ in the range $10^6-10^8$ GeV\cite{bs}. This is not high enough for thermal leptogenesis, 
which requires $T_R\agt 10^{10}$ GeV, but is high enough for non-thermal
leptogenesis, wherein heavy right handed neutrino states are produced via inflaton decay,
and which requires $T_R\agt 10^6$ GeV.

With the above spectrum, we expect an assortment of rich signals from gluino pair production
followed by 3-body gluino decays at the LHC\cite{so10lhc}. 
It is also possible that an axion might be detected
at direct axion search experiments. 
However, direct and indirect WIMP detection experiments should find a null result 
in the Yukawa-unified SUSY scenario with mixed axion/axino dark matter.


\begin{theacknowledgments}
I thank Dave Cline for organizing another fantastic Dark matter meeting.
I also thanks my various collaborators listed in the references who did all the work.
\end{theacknowledgments}


\end{document}